\documentclass[12pt,english]{article}
\usepackage{amsfonts,latexsym}
\usepackage{graphicx}
\usepackage{babel}

\title{Sagnac Effect of G\"odel's Universe}
\author{E. Kajari\thanks{email: endre.kajari@physik.uni-ulm.de}, R. Walser and W. P. Schleich\\
        Abteilung f\"ur Quantenphysik, Universit\"at Ulm, 89069 Ulm, Germany\vspace{0.5cm}\\
	        A. Delgado\\
        Departamento de F\'isica, Universidad de Concepci\'on,\\
	Casilla 160--C, Concepci\'on, Chile}

\date{\today}

\begin{document}

\maketitle

\begin{abstract}
We present exact expressions for the Sagnac effect of G\"odel's Universe. For this purpose we first derive
a formula for the Sagnac time delay along a circular path in the presence of an arbitrary stationary metric in
cylindrical coordinates. 
We then apply this result to G\"odel's metric for two different experimental 
situations: First, the light source and the detector are at rest relative to the matter generating the 
gravitational field. In this case we find an expression that is formally equivalent to the familiar nonrelativistic Sagnac 
time delay. Second, the light source and the detector are rotating relative to the matter. 
Here we show that for a special rotation rate of the detector the Sagnac time delay vanishes.
Finally we propose a formulation of the Sagnac time delay in terms of invariant physical quantities. We show that 
this result is very close to the analogous formula of the Sagnac time delay of a rotating coordinate 
system in Minkowski spacetime.
\end{abstract}

\newpage

\section{Introduction}
Phenomena in rotating coordinate systems have fascinated scientists since hundreds of years. Three examples may serve as an 
illustration of this statement: {\it i)}~The
Coriolis force is instrumental in the demonstration of the rotation of the earth with the help of the 
Foucault pendulum \cite{foucault}. {\it ii)} The Sagnac effect \cite{sagnac1,sagnac2,post,tartaglia,nandiK} measured by the 
Michelson-Gale-interferometer \cite{michelson} is the analogous optical tool. {\it iii)} Mach's principle 
\cite{weyl,isenbergwheeler,ciufoliniwheeler} ushers in a fresh view on 
the relativity of rotation. In the present paper we combine these three concepts and calculate the Sagnac effect of 
G\"odel's Universe. 

Einstein's theory of General Relativity \cite{ohanian} predicts gravito-magnetic forces
\cite{schleichscully,bonnor,laemmerzahl} due
to the rotation of massive objects. These forces give rise to the precession of nodal lines \cite{ciufolini} 
of two orbiting LAGEOS-satellites proving the first measurement of the Lense-Thirring
effect\cite{thirring1,thirring2,lense,mashoon}.
The newly emerging field of atom optics has opened a new arena \cite{hyper} for measuring the dipole character 
of the Lense-Thirring field with the help of an atom gyroscope \cite{gustavson2,gustavson,borde,borde2}. 

Rotation is not limited to a coordinate system or individual masses but even the universe can display
features of rotation. Indeed, in 1949 Kurt G\"odel \cite{goedel49,philos,goedel50} derived an exact solution of Einstein's field
equations, in which a homogeneous mass distribution rotates around every point in space. This solution shows
rather unusual properties \cite{hawking} such as closed time like world-lines.

In a recent paper \cite{delgado02} we have evaluated the Sagnac effect of G\"odel's Universe measured
in a laboratory size interferometer. For this purpose we have used the linearized G\"odel metric, which can be approximated
by the flat spacetime metric of a rotating coordinate system. 
In the present paper we calculate the  exact Sagnac effect for the G\"odel metric. 

\subsection{Why G\"odel's Metric?}

G\"odel's Universe has many fascinating features. Indeed, the inherent rotation of this Universe is only one interesting 
aspect. Even more intriguing is the lack of a global time ordering and the existence of closed time like world lines 
giving rise to the possibility of time travel. Such causal problems emerge also in other exact solutions of Einstein's
field equations, such as the Kerr metric. However, G\"odel's metric has the advantage that it is of rather compact form and most 
calculations can be carried out analytically.

It is commonly accepted that this model does not provide a matching description of our observed universe. 
Nevertheless, there is the emerging field of experimental cosmology in the laboratory \cite{coslab}. In particular, 
the examination of wave phenomena in curved spacetimes is a focus of research. For example, optical analogues
of black holes have been proposed by studying light propagation in moving media \cite{leonhardt} or sound propagation 
in condensed matter systems \cite{volovik,zoller}. In this context, it is important to ask 
how far one can extent such analogues in general. G\"odel's Universe can shed some light on this problem, 
since the existence of closed time like world lines curtails the expectation of a globally valid experimental 
analogue. Thus, it would be interesting to see, where and how the experimental realization ceases to exist.

Motivated by this idea we have initiated a study of wave propagation for the source free Maxwell equations 
in G\"odel's metric \cite{cohen,diplarbeit}. However, in the present 
paper we will focus on the Sagnac time delay within the limit of geometrical optics and draw the comparison 
to the Sagnac time delay in a rotating frame in flat Minkowski spacetime \cite{delgado02}.
In order to keep the paper self-contained 
and in view of the fact that this issue brings together researches from atomic, molecular and optical physics with experts in
General Relativity we have used this opportunity to combine our study of the Sagnac effect with a mini review of G\"odel's
Universe.

\subsection{Outline of the Paper}

Our paper is organized as follows: In Section \ref{sec1} we first briefly review the essential features of the Sagnac
effect and then derive an exact expression for the Sagnac time delay for a time independent metric in cylindrical
coordinates. We dedicate Section \ref{sec2} to a discussion of the G\"odel metric. We then apply in Section \ref{sagnactimedelay}
the expression of the Sagnac time delay to G\"odel's metric. In Section \ref{invform} we reformulate the Sagnac effect 
purely in terms of proper time delays of light pulses and compare it to the analogous expression for a rotating frame in 
flat Minkowski spacetime. We conclude by an outlook to possible laboratory 
realizations of the local light propagation in G\"odel's Universe.

\section{Sagnac Effect: Basics}\label{sec1}

In the present section we first recall the familiar Sagnac effect and then derive an exact expression for the Sagnac
time delay in terms of metric coefficients. We conclude by making contact with the familiar Sagnac effect using
the metric of a rotating coordinate system in flat Minkowski spacetime. 

\subsection{Familiar Sagnac Time Delay}

In 1913 George Sagnac performed an experiment \cite{sagnac1,sagnac2}, which he interpreted as an verification of the 
existence of the ether. In order to bring out the essential features of this phenomenon we consider two counter-propagating 
light pulses which
travel on a circle of radius $r_0$ due to an appropriate array of mirrors or a glass fiber. When the setup is 
at rest the two counter-propagating pulses arrive at the same time at the point of emission. However, when the 
arrangement rotates there is a time delay between them. This Sagnac time delay $\Delta t_S$ follows from an elementary, classical
argument \cite{chow} and reads
\begin{equation}
\Delta t_S\approx\frac{4\Omega_R}{c^2}\pi r_0^2.
\label{famsagnac}
\end{equation}
Hence, $\Delta t_S$ is proportional to the area $\pi r_0^2$ enclosed by the light beams and the rotation rate $\Omega_R$ of the ring 
with respect to the flat Minkowski spacetime. Moreover, we note that this result is an approximation and higher
order corrections in $\Omega_R$ will arise as discussed in Section \ref{timerotframe}.

The Sagnac effect is the basis of modern navigational systems. Moreover, it can also be generalized in
the framework of General Relativity \cite{schleichscully,scully}, as presented in the next section. 
In particular, the Sagnac effect is closely related to to the Hannay angle \cite{hannay} and 
the synchronization of clocks \cite{rosenblum} along a closed path.
 
\subsection{Time Delay for a Stationary Metric}

We now derive an exact expression for the Sagnac time delay $\Delta\tau_S$ in terms of the 
time independent metric coefficients $g_{\mu\nu}$ expressed in cylindrical coordinates
$x^\mu\equiv(x^0,x^1,x^2,x^3)\equiv(t,r,\phi,z)$. 
For a general treatment of the Sagnac effect of two counterpropagating 
light rays along any spatially closed path in an arbitrary metric we refer to \cite{bazanski2}. 
However, inhere we choose a particularly simple configuration and consider the propagation along a circle
of radius $r_0$ in the $z=z_0$ plane resulting in $dx^1=dr=0$ and $dx^3=dz=0$. For this situation the line element 
reads
\begin{equation}
ds^2=g_{\mu\nu}dx^\mu dx^\nu=g_{00}(dx^0)^2+2\,g_{02}\,dx^0dx^2+g_{22}(dx^2)^2,
\label{lightline}
\end{equation}
and vanishes for light $ds^2=0$. 

The angular velocity
\begin{equation}
\omega\equiv\frac{dx^2}{dx^0}
\label{angvel}
\end{equation}
of the light beam follows from the quadratic equation
$$g_{22}\,\omega^2+2\,g_{02}\,\omega+g_{00}=0,$$
giving rise to the two velocities
\begin{equation}
\omega_{\pm}=-\frac{g_{02}}{g_{22}}\pm\sqrt{\left(\frac{g_{02}}{g_{22}}\right)^2-\frac{g_{00}}{g_{22}}}.
\label{expressions}
\end{equation}
In order to get two solutions, which correspond to two ordinary counter-propagating light beams, we restrict 
ourselves to spacetime regions $G$, in which the above metric coefficients satisfy the conditions
\begin{equation}
g_{00}>0,\quad g_{22}<0\quad \forall\;\;x^\mu \in G.
\label{conditions}
\end{equation}
These conditions immediately imply that the angular velocity $\omega_{+}$ is positive and $\omega_{-}$ is negative for
all events in $G$, so that the monotony of the corresponding solutions $x^0_\pm(x^2)$ is guaranteed.
The condition $g_{00}>0$ is furthermore important to allow for an observer resting relative to the chosen spatial
coordinates and measuring the proper time, as discussed e.g. in \cite{landau}.

We can find the coordinate times $x^0_\pm$ of the first return of the counter-propagating light rays to their starting point
$x^2=\phi_0$
by integrating equation (\ref{angvel}) over one period $2\pi$ in positive and negative angular direction, respectively. 
Hence, we arrive at
\begin{equation}
x^0_\pm=\int\limits_{\phi_0}^{\phi_0\pm 2\pi}\frac{dx^2}{\omega_\pm}
=\pm\int\limits_0^{2\pi}\frac{dx^2}{\omega_\pm},
\nonumber
\end{equation}
where in the second step we have made use of the periodicity of the metric coefficients in the coordinate $x^2$.

When we recall the connection $d\tau=\sqrt{g_{00}(r_0,\phi_0,z_0)}\,dx^0/c$ between the coordinate time $x^0$ and the proper time 
$\tau$ measured by an observer resting at $(r_0,\phi_0,z_0)$, the corresponding proper times $\tau_\pm$ of the 
incoming light rays read
\begin{equation}
\tau_\pm=\pm\frac{1}{c}\sqrt{g_{00}(r_0,\phi_0,z_0)}\int\limits_0^{2\pi}\frac{dx^2}{\omega_\pm}.
\label{proptimes}
\end{equation}

The positive sign in $\tau_{+}$ corresponds to the light pulse propagating in the positive angular direction,
whereas the negative sign in $\tau_{-}$ denotes the proper time of the light pulse traveling in the negative 
angular direction.

The Sagnac proper time delay $\Delta\tau_S$ follows as
$$\Delta\tau_S\equiv\left(\tau_{+}-\tau_{-}\right)=\frac{1}{c}\sqrt{g_{00}(r_0,\phi_0,z_0)}\int\limits_0^{2\pi}
\frac{\omega_{+}+\omega_{-}}{\omega_{+}\omega_{-}}\,dx^2,$$
which reduces with help of the explicit expressions (\ref{expressions}) for $\omega_\pm$ to
\begin{equation}
\Delta\tau_S=-\frac{2}{c}\sqrt{g_{00}(r_0,\phi_0,z_0)}\int\limits_0^{2\pi}\frac{g_{02}}{g_{00}}\,dx^2.
\label{sagtimeint}
\end{equation}
If we further assume, that the metric coefficients $g_{02}$ and $g_{00}$ do not depend on the angular coordinate $x^2$, then 
the expression (\ref{sagtimeint}) for the Sagnac time delay reduces to the compact formula
\begin{equation}
\Delta\tau_S=-\frac{4\pi}{c}\;\frac{g_{02}}{\sqrt{g_{00}}}.
\label{sagtime}
\end{equation}
For negative values of $\Delta\tau_S$ the light pulse which propagates in the positive angular direction returns before
the other pulse and vice versa for positive $\Delta\tau_S$. 

We conclude by noting that the Sagnac time delay (\ref{sagtime}) is by construction only form-invariant under the special class
of coordinate transformations
$$x'^0=x'^0(x^0,x^1,x^3),\quad x'^1=x'^1(x^1,x^3),\quad x'^2=x^2,\quad x'^3=x'^3(x^1,x^3),$$
which change neither the frame of reference nor the angular coordinate.

\subsection{Time Delay in a Rotating Frame}\label{timerotframe}

We now want to apply formula (\ref{sagtime}) of the Sagnac time delay to the metric of a rotating coordinate 
frame in flat Minkowski spacetime. The line element in flat Minkowski spacetime reads in cylindrical 
coordinates $x^\mu=(t,r,\phi,z)$ 
\begin{equation}
ds^2=c^2dt^2-dr^2-r^2d\phi^2-dz^2.
\label{lineelflat}
\end{equation}
The coordinates $x'^\mu=(t',r',\phi',z')$ of a reference frame rotating with a rate $\Omega_R>0$ are defined by
the transformation equations
$$t\equiv t',\quad r\equiv r',\quad \phi\equiv\phi'+\Omega_R t',\quad z\equiv z',$$
and give rise to a line element $ds^2=g'_{\mu\nu}dx'^\mu dx'^\nu$ of the form
\begin{equation}
ds^2=(c^2-r'^2\Omega_R^2)\,dt'^2-dr'^2-r'^2d\phi'^2-dz'^2-2r'^2\Omega_R\,dt'd\phi'.
\label{lineelrot}
\end{equation}
To satisfy the conditions (\ref{conditions}) we have to restrict ourselves to the spacetime region
\begin{equation}
G'\equiv\left\{0\leq r'<\frac{c}{\Omega_R},\;0\leq \phi'<2\pi,\;-\infty<t',z'<\infty\right\}.
\label{rotsptimereg}
\end{equation}
When we now assume that the counter-propagating light rays travel along a circle with radius $r'_0$ and substitute
the metric coefficients $g'_{00}$ and $g'_{02}$ into (\ref{sagtime}), the Sagnac time delay in the rotating frame reads
\begin{equation}
\Delta\tau'_R=\frac{4\pi}{c^2}\,\frac{r'^2_0\Omega_R}{\sqrt{1-\left(\frac{r'_0 \Omega_R}{c}\right)^2}}.
\label{minsagtime}
\end{equation}
In the first approximation of $(r'\Omega_R/c)$ this expression reduces to the time delay of the familiar 
Sagnac effect (\ref{famsagnac}). 

However, also the limit of large rotation rates, where the square root in the denominator gets important is of 
interest. Indeed, in Section \ref{invform} we present an expression for the Sagnac time delay in G\"odel's Universe, which 
on first sight looks very different from (\ref{minsagtime}). Nevertheless, a closer view reveals that this formula
is numerically very close to the Minkowskian Sagnac time delay. 

\section{Essential Features of G\"odel's Universe}\label{sec2}

In order to gain some insight into the intricacies of G\"odel's Universe we briefly review two 
representations of G\"odel's metric that are convenient for our analysis of the Sagnac effect. 
Moreover, we sketch the conditions under which this metric solves Einstein's field equations. We then 
present tensorial quantities characterizing the time like velocity field of the matter 
generating the G\"odel metric. Furthermore, we make contact with 
the metric of flat spacetime in cylindrical coordinates and mention the symmetries and the causal structure 
of G\"odel's metric.
We conclude this section by summarizing special null geodesics, which will be important in expressing the Sagnac time 
delay by measurable quantities.

\subsection{Line Element and Einstein's Field Equations}

The line element $ds^2\equiv\bar{g}_{\mu\nu}\,d\bar{x}^\mu d\bar{x}^\nu$ given by G\"odel \cite{goedel49} 
in 1949 in dimensionless, cylindrical coordinates 
$\bar{x}^\mu\equiv(\bar{x}^0,\bar{x}^1,\bar{x}^2,\bar{x}^3)\equiv(\bar{t},\bar{r},\bar{\phi},\bar{z})$
with
\begin{equation}
G_G\equiv\left\{-\infty<\bar{t}<\infty,\: 0\le \bar{r}<\infty, \: 0\le \bar{\phi} <2\pi,\:
-\infty<\bar{z}<\infty\right\},
\label{altcoord}
\end{equation}
has the form
\begin{equation}
\frac{ds^2}{4a^2}=d\bar{t}^2-d\bar{r}^2-\left(\sinh^2\bar{r}-\sinh^4\bar{r}\right)\,d\bar{\phi}^2-d\bar{z}^2
+2\sqrt{2}\sinh^2\bar{r}\,d\bar{\phi}\,d\bar{t}.
\label{urmetric}
\end{equation}
The parameter $a>0$ has the unit of a length.

In the next section we derive a simple expression for the Sagnac effect. For this purpose, it is
convenient to use a slightly different form of G\"odel's metric, which can be obtained from (\ref{urmetric})
by the coordinate transformation 
\begin{equation}
t\equiv\frac{2a}{c}\bar{t},\quad r\equiv 2a \;\sinh\bar{r},\quad \phi\equiv\bar{\phi},\quad z\equiv 2a\bar{z}.
\label{coord}
\end{equation}
The resulting line element reads
\begin{equation}
ds^2=c^2dt^2-\frac{dr^2}{1+\left(\frac{r}{2a}\right)^2}
-r^2\left(1-\left(\frac{r}{2a}\right)^2\right)d\phi^2-dz^2+2r^2\frac{c}{\sqrt{2}a}dt\,d\phi\;. 
\label{metric}
\end{equation}
These new coordinates have now physical dimensions. 

We can convince ourselves that the metric coefficients 
\begin{equation}
(g_{\mu\nu})=\left(%
\begin{array}{c c c c}
c^2&0&r^2\frac{c}{\sqrt{2}a}&\phantom{-}0\vspace{2mm}\\
0&-\frac{1}{1+\left(\frac{r}{2a}\right)^2}&0&\phantom{-}0\\
r^2\frac{c}{\sqrt{2}a}&0&-r^2\left(1-\left(\frac{r}{2a}\right)^2\right)&\phantom{-}0\vspace{2mm}\\
0&0&0&-1
\end{array}%
\right)
\label{metriccoeff}
\end{equation}
corresponding to the line element (\ref{metric}) indeed solve Einstein's field equations 
$$R_{\mu\nu}-\frac{1}{2}g_{\mu\nu}R=\kappa T_{\mu\nu}+\Lambda g_{\mu\nu},\quad $$
by calculating explicitly the Ricci tensor $R_{\mu\nu}$ and the scalar curvature 
$R$, defined in Appendix \ref{convention}. Here $\Lambda$ denotes the cosmological constant and $\kappa\equiv(8\pi
G)/(c^4)$ with Newton's gravitational constant $G$ and the velocity of light $c$.

The energy momentum tensor $T_{\mu\nu}$ 
of an ideal fluid with mass density $\rho$ and pressure $p$ reads
$$T_{\mu\nu}\equiv\left(\rho+\frac{p}{c^2}\right)u_\mu u_\nu-p g_{\mu\nu}\;. $$
With the four-velocity $u^\mu=(1,0,0,0)$ of the matter generating the field, we find
$$R_{\mu\nu}=\frac{u_\mu u_\nu}{a^2 c^2},\quad R=\frac{1}{a^2}.$$
When we substitute these expressions into the field equations, we arrive at the 
two relations
$$\kappa\left(\rho+\frac{p}{c^2}\right)=\frac{1}{a^2c^2},\quad \kappa
p=\Lambda+\frac{1}{2a^2}\,,$$
which couple the length scale $a$ to the density $\rho$, the pressure $p$ and the cosmological
constant $\Lambda$.

\subsection{Time Like Velocity Field of the Ideal Fluid}

The essential properties of the motion of the ideal fluid generating the field
are characterized by the tensorial quantities $\theta$, $\sigma_{\alpha \beta}$ and
$\omega_{\alpha \beta}$ representing the volume expansion, the shear tensor and the rotation
tensor, respectively \cite{hawking,ehlers}. From the four-velocity $u^\mu\equiv(1,0,0,0)$
and the acceleration $a^\mu=u^\mu_{\;\, ;\nu}u^\nu=0$ of the congruence of time like curves 
belonging to the ideal fluid we find a vanishing volume expansion
$$\theta\equiv u^\mu_{\;\, ;\mu}=0,$$ 
and a vanishing shear tensor 
$$\sigma_{\alpha\beta}\equiv P^\mu_{\;\,\alpha}P^\nu_{\;\,\beta}\,u_{(\mu;\nu)}-\frac{1}{3}\theta
P_{\alpha\beta}=0.$$
Here we have introduced the projection tensor 
$$P^\alpha_{\;\,\beta}\equiv\delta^\alpha_\beta-\frac{1}{c^2}u^\alpha u_\beta$$
together with $u_{(\mu;\nu)}\equiv\left(u_{\mu;\nu}+u_{\nu;\mu}\right)/2$, where the semicolon
denotes the covariant derivative. 

However, we arrive at a non-vanishing rotation tensor
\begin{equation}
\omega_{\alpha\beta}\equiv P^\mu_{\;\,\alpha}P^\nu_{\;\,\beta}\,u_{[\mu;\nu]}=u_{[\alpha,\beta]}
=r\frac{c}{\sqrt{2}a}\left(\delta^2_\alpha\delta^1_\beta-\delta^2_\beta\delta^1_\alpha\right).
\label{rottensor}
\end{equation}
Here the square brackets are defined by $u_{[\mu;\nu]}\equiv\left(u_{\mu;\nu}-u_{\nu;\mu}\right)/2$
and the comma denotes the partial derivative. 

The corresponding rotation vector
\begin{equation}
\omega^\alpha\equiv\frac{1}{2}\,\varepsilon^{\alpha\beta\gamma\delta}\,u_\beta\,u_{\gamma;\delta}=\frac{c^2}{\sqrt{2}a}\delta^\alpha_3
\label{rotvektor}
\end{equation}
and the rotation scalar
\begin{equation}
\Omega_G\equiv\sqrt{\frac{1}{2}\,\omega^{\alpha\beta}\,\omega_{\alpha\beta}}=\frac{c}{\sqrt{2}a}>0.
\label{rotskalar}
\end{equation}
are constant in every point of G\"odel's Universe.

In the limit $a\rightarrow \infty$ all rotation 
quantities (\ref{rottensor}), (\ref{rotvektor}) and (\ref{rotskalar}) vanish. Moreover, for $r/(2a)\ll 1$ we find that 
the line element (\ref{metric}) can be approximated in first order by 
$$ds^2= c^2dt^2-dr^2-r^2d\phi^2-dz^2+2r^2\Omega_Gdt\,d\phi+\mathcal{O}(\Omega_G^2).$$
While the zeroth order approximation \cite{novello93} corresponds to the line element (\ref{lineelflat}) of flat spacetime 
in cylindrical coordinates the linear correction is reminiscent of the line element (\ref{lineelrot}) 
of a rotating coordinate frame in Minkowski spacetime. 

\subsection{Killing Vectors and Symmetry}

If a spacetime manifold with the metric $g_{\mu\nu}$ possesses symmetries, then they can be characterized by 
a special class of coordinate transformations $x'^\alpha=x'^\alpha(x^\beta)$ which satisfy the condition
\begin{equation}
g'_{\mu\nu}(x'^\alpha)=g_{\mu\nu}(x'^\alpha).
\label{iso}
\end{equation}
Such a coordinate transformation is called isometry. In particular, the infinitesimal isometries 
\begin{equation}
x'^\alpha=x^\alpha+\varepsilon\xi^\alpha(x^\beta)
\label{infiso}
\end{equation}
with $\varepsilon\ll 1$ are of special interest, since every continuous isometry can be constructed successively 
by these infinitesimal isometries. 

The infinitesimal transformation (\ref{infiso}) together with (\ref{iso}) and 
the transformation law of a metric yields the condition
\begin{equation}
\xi_{\alpha;\beta}+\xi_{\beta;\alpha}=0
\label{Killinggl}
\end{equation}
for the Killing vector field $\xi^\alpha(x^\beta)$. The solutions $\xi^\alpha$ of this linear system 
of partial differential equations characterize the symmetry of a given metric.

In the case of G\"odel's metric we find five Killing vectors as solutions of the Killing equation 
(\ref{Killinggl}). Three of them are immediately found from (\ref{iso}), since G\"odel's metric 
does not depend explicitly on the coordinates $(t,\phi,z)$. With the constants
$A,B,C,D,E$ the complete solution of (\ref{Killinggl}) reads
\begin{equation}
\xi^\alpha(t,r,\phi,z)=A\delta^\alpha_0+B\delta^\alpha_2+C\delta^\alpha_3+D\zeta^\alpha(r,\phi)
+E\zeta^\alpha\textstyle{\left(r,\phi-\frac{\pi}{2}\right)},
\label{killigoe}
\end{equation}
with
$$
\left( \begin{array}{c} \zeta^0\\ \zeta^1 \\ \zeta^2 \\ \zeta^3 \end{array} \right)\equiv
\frac{1}{\sqrt{1+\left(\frac{r}{2a}\right)^2}}\left( 
\begin{array}{c} \frac{r}{\sqrt{2}c}\cos\phi \\ a\left(1+\left(\frac{r}{2a}\right)^2\right)\sin\phi \\ 
\frac{a}{r}\left(1+2\left(\frac{r}{2a}\right)^2\right)\cos\phi\ \\ 0 \end{array} \right).
$$
\newline
Since (\ref{killigoe}) contains a time like Killing vector G\"odel's Universe is stationary. Moreover it is also 
spatially homogeneous. However, a more detailed analysis of the Killing vectors shows, that
G\"odel's metric is not static and also not isotropic. 
The latter feature is due to the existence of a rotational axis giving rise to a rotational symmetry 
in the $z=const.$ planes.

\subsection{Causal Structure}\label{causstruct}

\begin{figure}[h]
\begin{center}
\includegraphics[width=13.5cm]{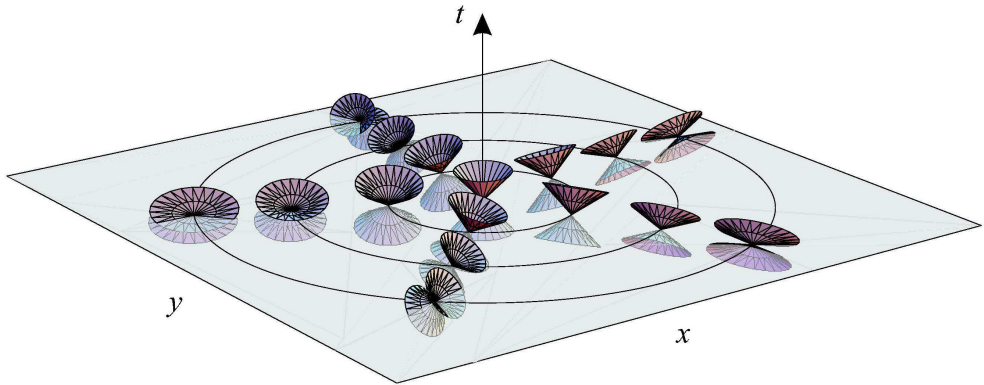}%
\caption{Light cones in G\"odel's metric represented in the $z=0$ plane. The middle circle of critical radius $r_G=2a$
separates the domains of different causal behavior. At every point of the inner domain the light cones lie outside 
of the $t=0$ plane as exemplified by the cones along the inner circle. In contrast, outside of the critical circle the 
cones cut through this plane.}%
\label{Lichtkegel}%
\end{center}
\end{figure}

The non-vanishing rotation scalar has a dramatic consequence for the causal structure of G\"odel's Universe.
In order to gain some insight into this feature, it is useful to consider infinitesimal light 
cones at different spatial points. Figure~\ref{Lichtkegel} depicts
such an arrangement.  The
cylindrical coordinates $(t,r,\phi,z)$ are embedded for illustration in a Cartesian frame 
\begin{equation}
(t,x\equiv r\cos\phi,y\equiv r\sin\phi,z)
\label{polarcoordem}
\end{equation}
and the third spatial coordinate $z$ is suppressed in the figure.
For every instant of time there exists a domain in space with closed time like world-lines. 
To visualize this property, we
consider the orientation of some selected light cones with respect to the plane of constant coordinate time $t$. 

At the origin of the coordinate system the axis of the light cone is orthogonal to the plane. As we move
away from the origin the light cones start to tilt, as indicated in Figure~\ref{Lichtkegel}. At the critical G\"odel radius 
$r_G=2a$, represented by the middle circle, the light cones are tangential to the plane of constant 
coordinate time $t$. 
This circle of radius $r_G$ is a light like curve. Outside this critical radius the inclination of the light cones 
increases further and allows the existence of closed time like curves, as shown by the outer circle in 
Figure~\ref{Lichtkegel}.

It is this peculiar feature of the causal structure which permits to connect two arbitrary events of spacetime 
by a time like curve, irrespectively of their ordering in the chosen coordinate time $t$. Indeed, we can start
from a point within the inner circle and cross the critical G\"odel radius to explore the world beyond this border.
During this trespassing we take a time like world line which spirals downwards into the past. Having regressed long 
enough on this trajectory we can finally return to the original causal domain, thus arriving before departing. 

We conclude by noting that closed time like world-lines are not limited to G\"odel's Universe.
They also appear in other exact solutions of Einstein's field equations for rotating mass
distributions. Examples include the rotating Kerr black hole \cite{kerr,carter} and the van Stockum 
rotating dust cylinder \cite{stockum}. Most recently is has been shown \cite{mallett}, that the 
gravitational field of a solenoid of light, that is a light beam bent along helical path also exhibits 
such closed time like world-lines.

\subsection{Null Geodesics}\label{nullgeo}

In the presence of a metric, the free motion of particles or the propagation of light rays  
is described by the geodesic equations
\begin{equation}
\frac{d^2x^\mu}{d\lambda^2}+\Gamma^\mu_{\;\alpha\beta}\frac{dx^\alpha}{d\lambda}\frac{dx^\beta}{d\lambda}
\equiv u^\mu_{\;\,;\nu}u^\nu=0
\label{geodesic}
\end{equation}
and by the condition 
$$g_{\mu\nu}u^\mu u^\nu=\epsilon^2.$$
For the definition of the Christoffel symbols $\Gamma^\mu_{\;\alpha\beta}$ we refer to Appendix~\ref{convention}
and $x^\mu(\lambda)=\big(t(\lambda),r(\lambda),\phi(\lambda),z(\lambda)\big)$ is the path of the particle or light beam. 
For massive particles $\epsilon$ denotes
the velocity $c$ of light and the curve parameter $\lambda$ represents its proper time. Moreover, the 
tangent vector $u^\mu\equiv dx^\mu/d\lambda$ is the four-velocity of the particle.
In the case of a light beam
we have to set $\epsilon=0$ and the curve parameter $\lambda$ has no physical meaning. 

The solutions of the geodesic equations for G\"odel's metric were first given by W. Kundt \cite{kundt}, 
independently  examined by S. Chandrasekhar and J. P. Wright \cite{chandra} and discussed in detail by Novello 
et. al. \cite{novellogeo}. 

In Section \ref{invform} we express the Sagnac time delay in measurable quantities only. For this 
purpose we need to know the geodesic motion of light for special initial conditions. 
Indeed, the light pulse shall start for the initial curve parameter $\lambda_0=0$ 
at the point $r(0)=0,\;z(0)=0$. Furthermore, the light ray shall have a vanishing $z$-component of the 
initial velocity, that is $u^3(0)=0$. Since we start at the origin, the radial velocity has to be positive, that is $u^1(0)>0$. 

In Appendix \ref{Intnullgeo} we outline a procedure for obtaining the general solution of the geodesic equations 
(\ref{geodesic}) for G\"odel's metric and integrate these equations subjected to these initial conditions. 

For the radial coordinate we find the expression
\begin{equation}
\frac{r(\lambda)}{2a}=\left|\sin\left(\textstyle{\frac{1}{2}}\eta\lambda\right)\right|,
\label{radius}
\end{equation}
where we have introduced the abbreviation 
\begin{equation}
\eta\equiv\sqrt{2}\,u^0(0)\Omega_G.
\label{etaform}
\end{equation}
Hence, when the curve parameter $\lambda$ reaches the value $\lambda_c\equiv(2\pi)/\eta$ the light pulse returns 
again to the origin $r=0$.

The coordinate time along these geodesics reads
\begin{equation}
t(\lambda)=-u^0(0)\,\lambda+\frac{\hspace{-1mm}2}{\Omega_G}
\left[\arctan\left(\sqrt{2}\tan\left(\eta\lambda/2\right)\right)+m(\lambda)\,\pi\right],
\label{time}
\end{equation}
where the integer
\begin{equation}
m(\lambda)\equiv\left[\frac{\eta \lambda}{2\pi}+\frac{1}{2}\right]_I
\label{mlambda}
\end{equation}
represents the greatest integer less than or equal to the number inside the brackets.

\begin{figure}[h]
\begin{center}
\includegraphics[width=11.5cm]{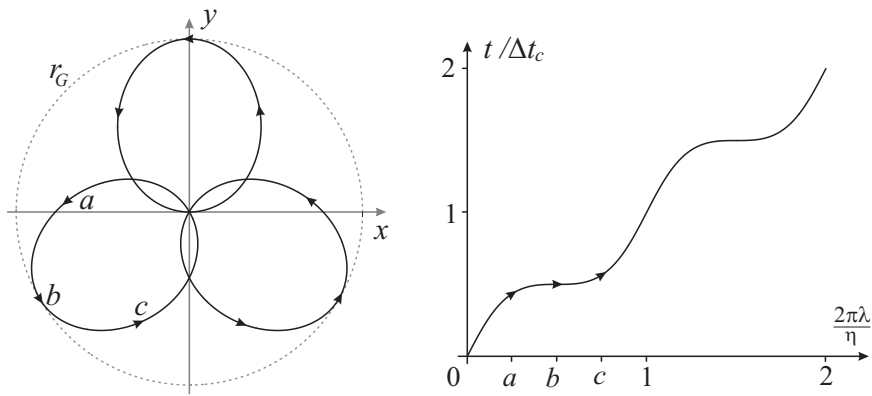}%
\caption{Null geodesics in the $(x,y)$--plane (left) and coordinate time $t(\lambda)$ extending over 2 cycles (right) 
for the special initial conditions $r(0)=0,\;u^3(0)=0$ and $\phi(0)=(0,\,2\pi/3,\,4\pi/3)$. The three values
$a,\,b,\,c$ of the curve parameter $\lambda$ mark the three different positions on a null geodesic on the left.}%
\label{geoplot}%
\end{center}
\end{figure}

Within one cycle, $0\leq \lambda\leq 2\pi/\eta\,$, the coordinate time $t(\lambda)$ increases by the time interval
$$\Delta t_c=\frac{\sqrt{2}\pi}{\Omega_G}\left(\sqrt{2}-1\right).$$
For the angle coordinate $\phi(\lambda)$ the integration yields
\begin{equation}
\phi(\lambda)=\phi(0)+\arctan\left(\sqrt{2}\tan\left(\eta\lambda/2\right)\right)+\left[m(\lambda)
-m\textstyle{\left(\lambda-\pi/\eta\right)}\right]\pi.
\label{phi}
\end{equation}
We conclude this section by illustrating the null geodesics for the special initial conditions 
in Figure \ref{geoplot}. 
The left side displays the geodesic 
motion in the $(x,y)$--plane embedded in the coordinate frame (\ref{polarcoordem}). 
The light signals emitted at the origin $r=0$ cycle in the positive angular direction. The right side of the figure shows the coordinate time $t(\lambda)$ for the increasing curve parameter $\lambda$. The inflection points in the time coordinate appear when the light ray touches the critical G\"odel radius.

\section{Sagnac Time Delay in G\"odel's Universe}\label{sagnactimedelay}

The goal of the present section is to derive an exact expression for the Sagnac time delay of
two counter-propagating light rays on a circle in the presence of G\"odel's metric.
We also analyze the Sagnac effect observed by a detector rotating relative to the ideal fluid. Finally we 
briefly discuss the dependence of the Sagnac time delay on the choice of the spatial coordinates.

\subsection{Sagnac Effect in the Rest Frame of the Ideal Fluid}\label{sagtimerestframe}

We start by considering a situation in which the emitter and the detector are at rest in the coordinate frame of the ideal 
fluid. The light pulses propagate along a circle of radius $r=r_0$ in the plane $z=z_0$.
Recalling G\"odel's metric (\ref{metriccoeff}) and the rotation scalar (\ref{rotskalar})
the relevant metric coefficients read
\begin{equation}
g_{00}=c^2,\quad g_{02}=r_0^2\Omega_G.
\label{koeffi}
\end{equation}
In order to satisfy the conditions (\ref{conditions}), we have to restrict ourselves to the spacetime region
$$G\equiv\left\{0\leq r <2a,\;0\leq \phi<2\pi,\;-\infty<t,z<\infty\right\}.$$ 
If we do not confine ourselves to this domain the coordinate time needed by one of the light pulses to return to the starting point will
be negative for radii larger than the critical G\"odel radius $r_G=2a$. This feature follows from the orientation
of the future light cones in Figure \ref{Lichtkegel}. 

Furthermore an observer resting with the detector at $r_0>r_G$ would measure
a negative proper time of the returning light ray, which travels backward in the coordinate time. By restricting our
experimental setup to the region $G$ we avoid such alien situations in which the light pulse returns before it
is emitted with respect to the chosen coordinate time.

When we substitute the coefficients (\ref{koeffi}) into formula (\ref{sagtime}) the Sagnac time delay for G\"odel's metric 
reads
\begin{equation}
\Delta\tau_S=-\frac{4\Omega_G}{c^2}\pi r_0^2.
\label{sagnactime}
\end{equation}

Since this expression is negative, we conclude from the definition of the proper time 
$\Delta\tau_S$, that the light pulse propagating in the positive angular direction will always return to its starting
point before the pulse propagating in the negative direction.

It is instructive to compare this result to the familiar Sagnac time delay $\Delta t_S$ given by (\ref{famsagnac}).
On first sight the absolute value of the Sagnac time delay (\ref{sagnactime}) in G\"odel's Universe seems to be 
identical to the familiar Sagnac effect (\ref{famsagnac}). However, when we recall that the coordinate $r_0$ in 
G\"odel's Universe does
not represent a proper distance,  we recognize that the similarity only arises from the special choice of our
spatial coordinates. We will return to this point in Section \ref{spatialcoorddep}.

We conclude by noting that the gravitational time delay in G\"odel's Universe has also been calculated in
\cite{ciumashoon}. However, the expression given in that paper is quadratic in the rotation scalar. 

\subsection{Sagnac Effect Measured by a Rotating Detector\label{rotatingdet}}

We now analyze the Sagnac time delay for a slightly different experimental situation. The light 
source emitting the two counter-propagating light rays and the detector are now no longer at rest relative to 
the ideal fluid, but rotate relative to it. We denote the corresponding rotation rate by $\Omega_D$. 
Due to this additional rotation the Sagnac time delay will also depend on $\Omega_D$. After providing an explicit 
expression for the Sagnac time delay for this situation we choose $\Omega_D$ such, that this time delay vanishes exactly.

We start from the metric (\ref{metriccoeff}) and perform the coordinate transformation 
$$t\equiv t',\quad r\equiv r',\quad \phi\equiv\phi'+\Omega_D t',\quad z\equiv z'$$
into the rotating frame of the detector.
In the new coordinates the line element takes the form
\begin{eqnarray*}
&&ds^2=\left(c^2-r'^2\Omega_D^2\left(1-\frac{r'^2}{4a^2}\right)+2r'^2\Omega_D\Omega_G\right) dt'^2
-\frac{dr'^2}{1+\left(\frac{r'}{2a}\right)^2}\\
&&-r'^2\left(1-\frac{r'^2}{4a^2}\right)d\phi'^2
-dz'^2+2r'^2\left(\Omega_G-\left(1-\frac{r'^2}{4a^2}\right)\Omega_D\right)dt'\,d\phi'\; .
\end{eqnarray*}
The conditions (\ref{conditions}) for the corresponding metric coefficients in the rotating frame lead to the 
restrictions
$$c^2-r'^2\Omega_D^2\left(1-\frac{r'^2}{4a^2}\right)+2r'^2\Omega_D\Omega_G>0$$
and 
$$0\leq r'<r_G$$
on the radial coordinate $r'$ and the rotation rate $\Omega_D$.
Indeed, these conditions lead to a nontrivial region $G$ of allowed radii $r'$ with respect to the chosen $\Omega_D$.

We again denote the radius of the circular light path by $r'_0$. When we insert the relevant coefficients 
$g'_{00}$ and $g'_{02}$ into formula (\ref{sagtime}) and take into account 
the rotation scalar (\ref{rotskalar}) we find the Sagnac time delay
$$\Delta\tau'_S=
-\frac{4\pi}{c}\frac{r'^2_0\left(\Omega_G-\left(1-\frac{r'^2_0}{4a^2}\right)\Omega_D\right)}
{\sqrt{c^2+r'^2_0\Omega_D\left(2\Omega_G-\left(1-\frac{r'^2_0}{4a^2}\right)\Omega_D\right)}}.$$

The special choice 
\begin{equation}
\Omega_D\equiv\frac{\Omega_G}{1-\frac{r'^2_0}{4a^2}}.
\end{equation}
of the rotation rate of the detector leads to a vanishing time delay $\Delta\tau'_S$ between the two counter-propagating 
light pulses. Hence, for infinitesimal radii $r'_0$, the rotation velocity $\Omega_D$
of the detector is equal to the rotation scalar $\Omega_G$ of G\"odel's Universe.

\subsection{A Different Choice of Spatial Coordinates}\label{spatialcoorddep}

We now return to the Sagnac time delay measured in the rest-frame of the ideal fluid of Section \ref{sagtimerestframe}.
According to equation (\ref{sagnactime}) the Sagnac time delay $\Delta\tau_S$ depends in a linear way on the rotation 
scalar $\Omega_G$ of G\"odel's Universe. 

We emphasize that this linearity is only due to the special choice of our space 
coordinates (\ref{coord}) in the line element (\ref{metric}). 
Indeed, we can also express the Sagnac time delay $\Delta\tau_S$ in the dimensionless space coordinates
of the original G\"odel line element (\ref{urmetric}), where the coordinate transformation (\ref{coord}) does not 
change the frame of reference. In these coordinates (\ref{altcoord}) the Sagnac time delay reads 
\begin{equation}
\Delta\tau_S=-\frac{8\pi}{\Omega_G}\sinh^2\bar{r}_0.
\label{sagtimenewcoord}
\end{equation}
Although in this representation the dependence of $\Delta\tau_S$ on $\Omega_G$ differs from that of 
equation (\ref{sagnactime}) both formulae contain the same physics. This invariance is hidden behind 
the explicit dependence of $\Delta\tau_S$ in (\ref{sagnactime}) and (\ref{sagtimenewcoord}) on the spatial coordinates, 
which have no immediate physical meaning. In the next section we are going to rectify this problem by
introducing an invariant formulation via an operational definition of the radial coordinate. 

\section{Invariant Formulation of the Sagnac Time Delay}\label{invform}

The different formulae (\ref{sagnactime}) and (\ref{sagtimenewcoord}) for the Sagnac time delay bring out most 
clearly the question: How can we reformulate the radial coordinate of the mirrors in a measurable quantity? 

In the present section we outline a measurement strategy to address this question and then reformulate the Sagnac 
time delay solely in terms of proper times of light signals. The obtained relation allows for an interesting 
comparison to expression (\ref{minsagtime}) for the Sagnac time delay in a rotating frame of reference.

\subsection{Operational Definition of the Radial Coordinate}\label{replacesag}

All expressions for the Sagnac time delay derived so far contain a radial coordinate. What is a measurement strategy 
for obtaining the radius $r_0$ of the circular path of the light pulses? 

\begin{figure}[h]
\begin{center}
\includegraphics[width=11.5cm]{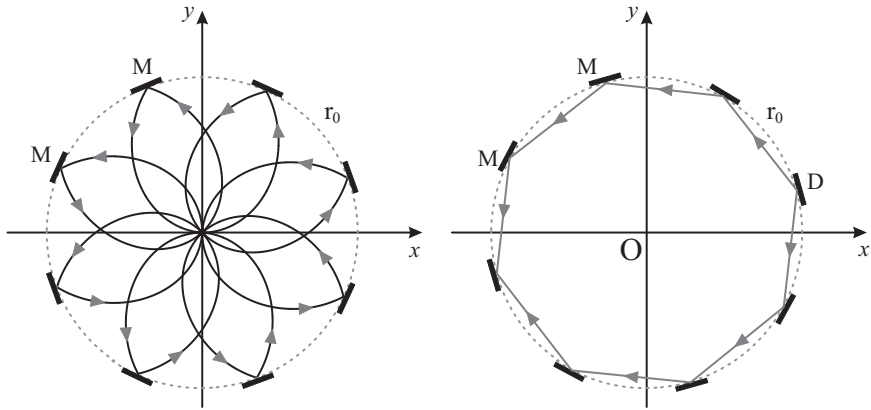}%
\caption{Illustration of the measurement procedure for the radius $r_0$ of a circle (left) and of the Sagnac time delay
$\Delta\tau_S$ (right). Light rays start at the origin and propagate on
null geodesics to the mirrors (M), where they are reflected back to the origin (left). The proper time $\Delta\tau_M$ 
between emission and arrival of the light pulses is measured by an observer resting at the origin. The figure on the right 
illustrates the typical experimental arrangement for the Sagnac effect with a detector (D) on the circle.
The straight lines between two consecutive mirrors are an approximation to the corresponding null geodesics.
However, the curves shown on the left are exact null geodesics (see Appendix \ref{Intnullgeo}).}%
\label{Lichtbahnen}%
\end{center}
\end{figure}

We answer this question in the spirit of 
\cite{ohanian,wheelermarzke,bodenner}. We send light signals along their null geodesics from the center of the circle 
to every mirror $M$ in the $z=0$ plane, as illustrated on the left side of Figure \ref{Lichtbahnen}. 
The mirrors are arranged in such a way as to reflect the light back to the point of emission. 
Moreover they are lined up in the $z=0$ plane as to ensure that the pulses emitted simultaneously at the center and reflected from the mirrors all return to the center at the same time.

Hence, the radius $r_0$ can be expressed by the proper time $\Delta\tau_M$ 
between the emission and detection of the reflected pulses measured by an observer resting at the origin 
$r=0$. 
Once the circular setup of mirrors has been established by this operational procedure the mirrors have to be 
readjusted to guide the counter-propagating light rays in this circular Sagnac configuration 
with radius $r_0$, as indicated in the right of Figure \ref{Lichtbahnen}.

We conclude by emphasizing that our experimental implementation is only the simplest conceivable model. It is
a gedanken experiment, which even requires an infinite amount of mirrors. However, a practical realization 
has many caveats which need to be considered, such as the finite number of mirrors, the propagation of the light 
pulses on null geodesics in-between and the applicability of geometrical optics. 
In fact, the case of a finite number of mirrors has been carried out in principle 
by \cite{bazanski2,bazanski1}, 
where the Sagnac time delay in a rotating frame in Minkowski spacetime serves as an example of the given method. 

\subsection{Invariant Formulation}

We are now in a position to establish a connection between the Sagnac time delay $\Delta\tau_S$ 
and the rotation scalar $\Omega_G$ in terms of the proper time interval $\Delta\tau_M$, measured by an observer resting 
at the origin $r=0$ of the coordinate system. We start by calculating the coordinate time for the light pulses to 
propagate to the mirrors and back. For this purpose
we recall (\ref{radius}) -- (\ref{time}). We then translate this time interval into proper
time and express the radius $r_0$ by the Sagnac time delay making use of equation (\ref{sagnactime}).

Since our mirrors are positioned at the radius $r_0<r_G$ the curve parameter $\lambda_0$ corresponding to this radius  
can be found from equation (\ref{radius}). Within a period $0\leq\eta\lambda<2\pi$ two values of
$\lambda$ correspond to $r_0$ and we choose the first solution $\lambda_0$ within the 
interval $0<\eta\lambda_0<\pi$.
The coordinate time between the emission of a light pulse at $r=0$ and the reflection of it from $r=r_0$ follows from
(\ref{time}) and (\ref{mlambda}) and reads
\begin{equation}
t(\lambda_0)=-u^0(0)\,\lambda_0+\frac{\hspace{-1mm}2}{\Omega_G}
\arctan\left(\sqrt{2}\tan\left(\eta\lambda_0/2\right)\right).
\label{rechnung2}
\end{equation}
Since G\"odel's metric is stationary, spatially homogeneous and possesses rotational symmetry in the planes 
$z={\rm const.}$, it takes twice the time $t(\lambda_0)$ to travel to and from the mirrors. 
For an observer resting at the origin $r=0$ the proper time $\Delta\tau_M$ between emission and arrival of the 
light pulses is given by $ds=c\,d\tau=c\,dt$, and consequently 
$$\Delta\tau_M=2t(\lambda_0).$$
When we use equation (\ref{radius}) for $\lambda_0$ in (\ref{rechnung2}) we arrive at the expression
\begin{equation}
\Omega_G\Delta\tau_M=-2\sqrt{2}\arcsin\left(\frac{r_0}{2a}\right)+4
\arctan\left(\frac{\sqrt{2}\left(\frac{r_0}{2a}\right)}{\sqrt{1-\left(\frac{r_0}{2a}\right)^2}}\right)
\label{beoeigenzeit}
\end{equation}
for the proper time $\Delta\tau_M$. Here we have recalled the definition (\ref{etaform}).

In order to make the connection to the absolute value of the Sagnac time delay $\Delta\tau_S$, we recall 
from (\ref{sagnactime}) the relation
$$|\Delta\tau_S|=\frac{4\Omega_G}{c^2}\pi r_0^2=\frac{8\pi}{\Omega_G}\left(\frac{r_0}{2a}\right)^2,$$
keeping in mind, that the negative value of $\Delta\tau_S$ is only due to a faster propagation of the light pulse in the 
positive angular direction.

This formula allows us to express the radius $r_0$ in (\ref{beoeigenzeit}) in terms of the absolute value $|\Delta\tau_S|$
of the Sagnac time delay. In order to avoid the appearance of the square root of $\Delta\tau_S$ 
it is convenient to introduce the dimensionless parameters
\begin{equation}
\Delta_S^2\equiv\frac{1}{8\pi}\Omega_G|\Delta\tau_S|=\left(\frac{r_0}{2a}\right)^2
\label{parameterS}
\end{equation}
with $0<\Delta_S<1$ and
\begin{equation}
\Delta_M^2\equiv\frac{1}{8\pi}\Omega_G\Delta\tau_M.
\label{parameterM}
\end{equation}
When we replace the proper time $\Delta\tau_M$ and the radius $r_0$ in (\ref{beoeigenzeit}) by their corresponding 
dimensionless parameters (\ref{parameterS}) and (\ref{parameterM}) we arrive at the transcendental equation
\begin{equation}
4\pi\Delta_M^2=2\arctan\left(\frac{\sqrt{2}\Delta_S}{\sqrt{1-\Delta_S^2}}\right)-\sqrt{2}\arcsin{\Delta_S}.
\label{connectexact}
\end{equation}
This relation is quite a remarkable result since it provides an invariant formulation of the Sagnac time delay 
in G\"odel's Universe. Indeed, for a given value of the $\Delta_M$, that is for the radius of the circular path measured in propagation time $\Delta\tau_M$ of light, the solution of this equation is the scaled Sagnac time delay $\Delta_S$. Since we have scaled all proper times in terms of the rotation scalar the transcendental equation contains no parameters of G\"odel's Universe. The information about the metric reflects itself solely in the form of the equation.

\subsection{Comparison to the Rotating Frame in Flat Spacetime}\label{compar}

We now compare the transcendental equation (\ref{connectexact}) for the Sagnac time delay of G\"odel's 
metric with the corresponding formula for the time delay (\ref{minsagtime}) found in a rotating frame 
of reference in Minkowski spacetime. 

The first step consists of replacing the radial coordinate in equation (\ref{minsagtime}) by a measurable time. For this purpose we apply the same measurement strategy to the rotating frame as discussed in \ref{replacesag}. We then substitute the radial coordinate $r'_0$ by the proper time interval $\Delta\tau'_M$ of a light ray, which returns after reflection from the circular mirror arrangement at $r'_0$ to the origin. Since the observer at the origin is at rest relative to the inertial frame of Minkowski spacetime, we find immediately the proper time 
$$\Delta\tau'_M=\frac{2r'_0}{c}.$$
For the comparison of the Sagnac time delay (\ref{minsagtime}) with the expression (\ref{connectexact}) 
it is useful to introduce the dimensionless parameters 
\begin{equation}
\Delta'^{2}_R\equiv\frac{1}{8\pi}\Omega_R\Delta\tau'_R
\label{parameterRs}
\end{equation}
and
\begin{equation}
\Delta'^{2}_M\equiv\frac{1}{8\pi}\Omega_R\Delta\tau'_M,
\label{parameterMs}
\end{equation}
in complete analogy to the formulae (\ref{parameterS}) and (\ref{parameterM}).

Using these parameters we can cast the Sagnac time delay (\ref{minsagtime}) into the form 
$$2\Delta'^2_R=\frac{\left(4\pi\Delta'^2_M\right)^2}{\sqrt{1-\left(4\pi\Delta'^2_M\right)^2}},$$
which in contrast to (\ref{connectexact}) is an explicit formula for $\Delta'_R$ in terms of $\Delta'_M$.

In order to obtain a formula analogous to (\ref{connectexact}) we solve this equation for $4\pi\Delta'^2_M$, 
which yields 
\begin{equation}
4\pi\Delta'^2_M=\sqrt{2}\Delta'_R\sqrt{\sqrt{1+\Delta'^4_R}-\Delta'^2_R}.
\label{connectrot}
\end{equation} 
Due to our restriction (\ref{rotsptimereg}) to the spacetime region $G'$ the parameter $\Delta'_M$ has to 
satisfy the condition
$$0\leq 4\pi \Delta'^2_M<1.$$

\begin{figure}[h]
\begin{center}
\hspace{-0.2cm}
\includegraphics[width=8.7cm]{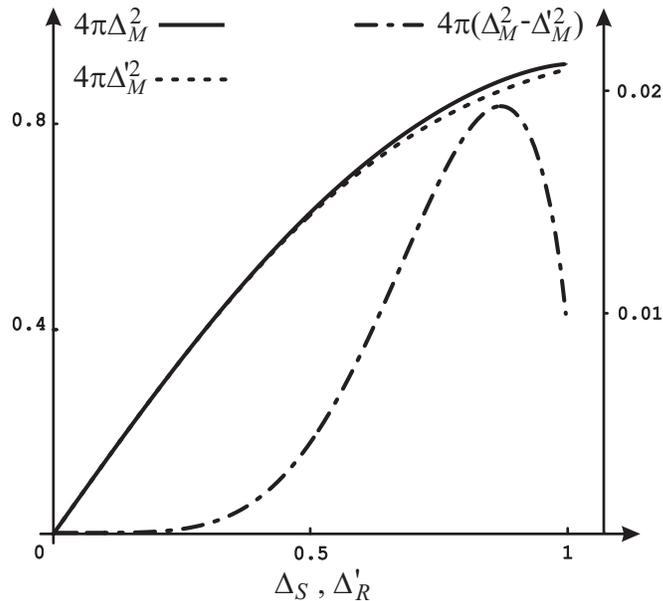}%
\caption{Comparison between the Sagnac time delay for G\"odel's Universe (solid line) and for a rotating coordinate system
in flat spacetime (dotted curve). Scaled proper times $4\pi\Delta_M^2$ and $4\pi\Delta'^2_M$ of a light signal 
propagating from the origin to a mirror on the circle and back in G\"odel's Universe and in a rotating frame of reference 
in flat spacetime, plotted versus the Sagnac time delays $\Delta_S$ and $\Delta'_R$ in scaled units. 
While the left ordinate applies to the absolute values of those scaled time delays, the dashed-dotted line magnifies 
the residual differences $4\pi(\Delta'^2_M-\Delta_M^2)$ as indicated by the right ordinate.}%
\label{invsagnac}%
\end{center}
\end{figure}

Figure \ref{invsagnac} illustrates the differences in the curves defined by (\ref{connectexact}) and 
(\ref{connectrot}). We find that for our measurement strategy the difference between the solid and dotted curve is very small. This is surprising, as the underlying physical systems with G\"odel's
Universe on the one hand and the rotating frame in flat Minkowski spacetime on the other hand, differ substantially in their global behavior. 

This similarity can also be understood from the Taylor series expansion of (\ref{connectexact}) and 
(\ref{connectrot}). Indeed, for small values of $\Delta_S$ or $\Delta'_R$ we find  
\begin{equation}
4\pi\Delta_M^2=\sqrt{2}\left(\Delta_S-\frac{1}{2}\Delta_S^3+\frac{11}{40}\Delta_S^5+\mathcal{O}(\Delta_S^7)\right),
\label{appform}
\end{equation}
and 
\begin{equation}
4\pi\Delta'^2_M=\sqrt{2}\left(\Delta'_R-\frac{1}{2}\Delta'^3_R+\frac{1}{8}\Delta'^5_R+\mathcal{O}(\Delta'^7_R)\right).
\end{equation}

Thus the two equations determining the Sagnac time delay only differ in the contribution of the fifth power 
of $\Delta_S$ or $\Delta'_R$.

\section{Summary and Outlook}

In conclusion we have investigated the Sagac time delay in G\"odel's Universe. Our analysis generalizes the results of our
previous examination \cite{delgado02}, which were limited to small radial dimensions and rotation rates, to 
arbitrary sizes of the circular Sagnac interferometer and arbitrary rotation rates of G\"odel's Universe. The previous results exhibited a close relation between the Sagnac effect in G\"odel's Universe and in a rotating frame in flat spacetime. By obtaining exact and invariant expressions for the Sagnac time delays in both systems, valid for arbitrary rotation rates and radii within $G$ and $G'$, we have demonstrated inhere, that this statement is also valid in general.

The original G\"odel solution is not in agreement with cosmological observations of the background
radiation \cite{bunn}. Therefore, one might be tempted to dismiss this remarkable exact solution of Einstein's field equations
altogether. However, recently analogies between light propagation
in curved spacetimes and optics in moving media \cite{leonhardt} or sonic propagation in condensed matter systems 
\cite{volovik,zoller} have instigated renewed interest in this subject. For example, these articles consider  
experimental analogues of a gravitational black hole. Here, the event horizon emerges when the flow velocities of the 
optical medium or the condensate exceed the velocities of light or sound in these media, respectively.

How far can we push these analogies? Can we find an optical or sonic realization of the light propagation in G\"odel's
Universe? Unfortunately, a global simulation of the G\"odel spacetime via an experimental analogue seems not to be 
very probable due to the causal structure of this metric discussed in Section \ref{causstruct}. 
But since we can approximate the local light propagation in G\"odel's Universe by a rotating coordinate frame 
in Minkowski spacetime it should be possible to translate the physics of wave propagation in curved space 
locally to cold quantum gases.
Vortices \cite{holland,fetter,nandi} or alternatively Abrikosov lattices \cite{ketterle} might offer a possible route 
to an analogue of G\"odel's Universe in the laboratory.

\section{Acknowledgments}
We thank I. Ciufolini, J. Ehlers, G. Sch\"afer and G. S\"ussmann and A. Wolf for many fruitful discussions. 
Moreover, we have thoroughly 
enjoyed the stimulating atmosphere at the Hyper meeting in Paris and appreciated the financial support, which made
our participation possible. We are also very grateful to C. L\"ammerzahl for his editorial efforts.

\appendix

\section{Definitions and Conventions\label{convention}}

Throughout this article we use the signature $(+,-,-,-)$ of our metric $g_{\mu\nu}$ with the determinant
$g\equiv\det{g_{\mu\nu}}$
and denote the covariant derivative with a semicolon and the ordinary partial derivative with a comma.

The antisymmetric tensor
$$\varepsilon^{\alpha\beta\gamma\delta}\equiv\frac{1}{\sqrt{-g}}\Delta^{\alpha\beta\gamma\delta}$$ 
is defined in terms of the Levi-Cevita-Symbol
\begin{equation}
\Delta^{\alpha\beta\gamma\delta}\equiv\left\{\begin{array}{l} \phantom{-}1\quad \mbox{for an even permutation}\\
-1\quad \mbox{for an odd permutation}\\
\phantom{-}0\quad\mbox{otherwise}\end{array}\right\}.
\nonumber
\end{equation}
The components of the Riemann tensor
$$R^\mu_{\;\,
\alpha\beta\gamma}\equiv\Gamma^\mu_{\;\alpha\gamma,\beta}-\Gamma^\mu_{\;\alpha\beta,\gamma}
+\Gamma^\mu_{\;\rho\beta}\Gamma^\rho_{\;\alpha\gamma}-\Gamma^\mu_{\;\rho\gamma}\Gamma^\rho_{\;\alpha\beta}$$
result from the Christoffel symbols
$$\Gamma^\mu_{\;\alpha\beta}\equiv\frac{1}{2}\,g^{\mu\nu}\left(g_{\nu\alpha,\beta}+g_{\nu\beta,\alpha}-g_{\alpha\beta,\nu}\right).$$
The Ricci tensor and the scalar curvature
$$ R_{\alpha\beta}\equiv R^\mu_{\;\,\alpha\mu\beta}\;,\quad R\equiv R^\mu_{\;\,\mu}$$
follow by contraction. 

The covariant derivative of a contravariant vector field $T^\alpha$ reads
$$T^\alpha_{\,\;;\beta}\equiv T^\alpha_{\;\,,\beta}+\Gamma^\alpha_{\;\mu\beta}T^\mu$$
whereas for a covariant vector field it takes the form
$$T_{\alpha;\beta}\equiv T_{\alpha,\beta}-\Gamma^\mu_{\;\alpha\beta}T_\mu.$$

\section{Integration of Null Geodesics}\label{Intnullgeo}

In this Appendix we outline a procedure to obtain the solutions of the geodesic equations in G\"odel's metric.
Since the main goal of the paper is the Sagnac time delay and in particular, an invariant formulation of it, 
we confine ourselves to special initial conditions appropriate for our experimental setup.

\subsection{General Idea}

The central idea for solving the geodesic equations relies on finding simple expressions for the constants of motion. 
Some of these constants can easily be obtained if the metric possesses Killing vectors $\xi^\mu$. 
This feature can be understood by contracting the geodesic equations (\ref{geodesic}) with the contravariant components 
of a Killing vector, that is
$$\xi_\mu u^\mu_{\;\,;\nu}u^\nu=\left(\xi_\mu u^\mu\right)_{;\nu} u^\nu-\xi_{\mu;\nu}u^\mu u^\nu=0.$$
When we recall the Killing equation (\ref{Killinggl}) and note, that the covariant derivative of a scalar is just the 
ordinary partial derivative, we find
$$\frac{d}{d\lambda}\left(\xi_\mu u^\mu\right)=0.$$
Hence, every Killing vector corresponds to a constant of motion. 

In addition the norm of the tangent vector 
$u^\mu(\lambda)$ is constant along the geodesics and therefore yields another constant of motion
\begin{equation}
g_{\mu\nu} u^\mu u^\nu=\epsilon^2.
\label{norm}
\end{equation}

Since the metric coefficients (\ref{metriccoeff}) depend only on the radial coordinate $r$, the solution of the geodesic
equations (\ref{geodesic}) can be found by making use of the first three Killing vectors in (\ref{killigoe}) to 
obtain the simple constants of motion 
\begin{eqnarray}
&&\hspace{-1cm}A=u_0(\lambda)=c^2 u^0(\lambda_0)+r^2(\lambda_0)\,\Omega_G u^2(\lambda_0)\label{const1}\\
&&\hspace{-1cm}B=u_2(\lambda)=r^2(\lambda_0)\,\Omega_G u^0(\lambda_0)-r^2(\lambda_0)
\left(1-\left(\frac{r(\lambda_0)}{2a}\right)^2\right) u^2(\lambda_0)\label{const2}\\
&&\hspace{-1cm}C=u_3(\lambda)=-u^3(\lambda_0)\label{const3}.
\end{eqnarray}
Here $\lambda_0$ denotes the initial curve parameter and in the second step of these equations we have made 
use of the relation $u_\mu(\lambda_0)=g_{\mu\nu}(\lambda_0)u^\nu(\lambda_0)$ with the metric 
coefficients (\ref{metriccoeff}).

We substitute the expressions (\ref{const1})-(\ref{const3}) for the covariant components of the four--velocity $u^\mu$ into 
the constant of motion (\ref{norm}) and find with the contravariant metric coefficients 
\begin{equation}
(g^{\mu\nu})=\left(%
\begin{array}{c c c c}
\frac{1}{c^2}\,\frac{1-\left(\frac{r}{2a}\right)^2}{1+\left(\frac{r}{2a}\right)^2}&0&\frac{\Omega_G}{c^2}\,\frac{1}{1+\left(\frac{r}{2a}\right)^2}&\phantom{-}0\vspace{1mm}\\
0&-\left(1+\left(\frac{r}{2a}\right)^2\right)&0&\phantom{-}0\vspace{1mm}\\
\frac{\Omega_G}{c^2}\,\frac{1}{1+\left(\frac{r}{2a}\right)^2}&0&-\frac{1}{r^2}\,\frac{1}{1+\left(\frac{r}{2a}\right)^2}&\phantom{-}0\vspace{2mm}\\
0&0&0&-1
\end{array}%
\right).
\label{contrametriccoeff}
\end{equation}
the equation
\begin{eqnarray}
g^{\mu\nu}u_\mu u_\nu=&&\frac{A^2}{c^2}\,\frac{1-\left(\frac{r}{2a}\right)^2}{1+\left(\frac{r}{2a}\right)^2}
+\frac{\Omega_G}{c^2}\,\frac{2AB}{1+\left(\frac{r}{2a}\right)^2}-\frac{B^2}{r^2}\,\frac{1}{1+\left(\frac{r}{2a}\right)^2}
-C^2\nonumber\\
&&-\left(1+\left(\frac{r}{2a}\right)^2\right)(u_1(\lambda))^2=\epsilon^2.\label{radeqn}
\end{eqnarray}

Inserting $u_1=g_{11}u^1$ into this expression yields a differential equation in the radial coordinate $r$. The solution $r(\lambda)$ 
can then be used to formulate the differential equations for the coordinates $t$ and $\phi$ by making use of the relations
$u^0=g^{0\mu}\,u_\mu$ and $u^2=g^{2\mu}\,u_\mu$. The solution of the third spatial coordinate $z$ can be obtained 
from the last constant of motion (\ref{const3}) 
\begin{equation}
u^3=g^{33}\,u_3=-C,\quad z(\lambda)=u^3(\lambda_0)(\lambda-\lambda_0)+z(\lambda_0).
\label{zcoord}
\end{equation}
Thus, the motion in $z$--direction is like a free motion in one dimension in flat Minkowski spacetime. 

\subsection{Special Initial Condition}

In the remainder of this Appendix we focus on the solution of the geodesic equations for light, 
that is $\epsilon=0$ in (\ref{norm}), subjected to the special initial conditions associated with
the experimental arrangement considered in Section \ref{invform}. 
In particular, the light pulse starts with the initial curve 
parameter $\lambda_0=0$ at the position $r(0)=0,\;z(0)=0$ and propagates in the $z=0$ plane. Therefore, we have to choose
$u^3(0)=0$. We emphasize, that for $r=0$ the angle velocity $u^2(0)$ looses its meaning. Moreover, since we start at the origin
and $r$ is positive, we have to take $u^1(0)>0$. 

These special initial conditions reduce the constants of motion (\ref{const1})--(\ref{const3}) to 
\begin{eqnarray}
&&A=u_0(\lambda)=c^2 u^0(0)>0 \label{cons1}\\
&&B=u_2(\lambda)=0 \label{cons2}\\
&&C=u_3(\lambda)=0.\label{cons3}
\end{eqnarray}
These compact expressions allow us to integrate the differential equation (\ref{radeqn}) 
for our radial coordinate $r(\lambda)$ as shown in the next section.

\subsection{Radial Coordinate}

The constants of motion (\ref{cons1})--(\ref{cons3}) reduce the differential equation (\ref{radeqn}) to
$$c^2\left(1-\left(\frac{r}{2a}\right)^2\right)(u^0(0))^2-\left(1+\left(\frac{r}{2a}\right)^2\right)^2(u_1(\lambda))^2
=0,$$
which leads with $u_1=g_{11}\,u^1$ and $u^1(0)>0$ to
\begin{equation}
\frac{dr}{d\lambda}=\pm c\,u^0(0) \sqrt{1-\left(\frac{r}{2a}\right)^2}.
\label{DGLr}
\end{equation}
The two equations correspond to the two different signs of the radial velocity.

Separation of variables 
$$\int\limits_0^{r(\lambda)}\frac{dr}{\sqrt{1-\left(\frac{r}{2a}\right)^2}}=\pm c\,u^0(0)\int\limits_0^\lambda d\lambda$$ 
finally yields the radial geodesic
\begin{equation}
\frac{r(\lambda)}{2a}=\left|\sin\left(\textstyle{\frac{1}{2}}\eta\lambda\right)\right|.
\label{radius2}
\end{equation}
Here we have introduced the periodicity rate
\begin{equation}
\eta\equiv\sqrt{2}\,u^0(0)\Omega_G
\label{perrate}
\end{equation}
in such a way, that one cycle of the radial coordinate $r(\lambda)$ corresponds to the curve parameter
$\lambda=2\pi/\eta$. 

It is apparent, that the solution (\ref{radius2}) is not differentiable at $r=0$ . 
This feature is due to the transformation to polar coordinates, which is singular at $r=0$. 
Furthermore we recognize that these null geodesics touch the critical G\"odel radius $r_G=2a$ for $\eta\lambda=\pi$.

\subsection{Coordinate Time}
The explicit expression (\ref{radius2}) for our radial coordinate allows us now to find the corresponding expressions for
both the coordinate time $t(\lambda)$ and the angular coordinate $\phi(\lambda)$. For this purpose
we use the relation $u^0=g^{0\mu}u_\mu$ 
with the contravariant metric coefficients (\ref{contrametriccoeff}) and the constants
of motion (\ref{cons1}) and (\ref{cons2}) and arrive at
$$u^0(\lambda)=u^0(0)\,\frac{1-\left(\frac{r(\lambda)}{2a}\right)^2}{1+\left(\frac{r(\lambda)}{2a}\right)^2}\geq 0.$$
Hence, the coordinate time is increasing monoto\-nously.

Substitution of (\ref{radius2}) into the above equation leads to 
\begin{equation}
t(\lambda)=-u^0(0)\,\lambda+2u^0(0)\int\limits_0^\lambda \frac{d\lambda}{1+\sin^2\left(\frac{1}{2}
\eta\lambda\right)},
\label{rechnung1}
\end{equation}
which after integration yields
\begin{equation}
t(\lambda)=-u^0(0)\,\lambda+\frac{\hspace{-1mm}2}{\Omega_G}
\left(\arctan\left(\sqrt{2}\tan\left(\eta\lambda/2\right)\right)+m(\lambda)\,\pi\right).
\label{time2}
\end{equation}
Here we have used (\ref{perrate}) and introduced the integer
\begin{equation}
m(\lambda)\equiv\left[\frac{\eta \lambda}{2\pi}+\frac{1}{2}\right]_I
\label{gausschekl}
\end{equation}
where the brackets denote the greatest integer less than or equal to the number inside them.
The integer $m$ guarantees the continuity, differentiability and monotony of $t(\lambda)$ for arbitrary
$\lambda \geq 0$.

\subsection{Angular Coordinate}

The dependence of the angular coordinate $\phi(\lambda)$ on the curve parameter can be found in a way similar to
the one of the coordinate time $t(\lambda)$. With the help of the relation $u^2=g^{2\mu}u_\mu$ and the constants of
motion (\ref{cons1}) and (\ref{cons2}) we arrive at
$$u^2(\lambda)=\frac{u^0(0)\,\Omega_G}{1+\left(\frac{r(\lambda)}{2a}\right)^2}.$$
Since the angular velocity $u^2$ looses its meaning at the point $r=0$ we perform the integration over
one cycle only, that is $0\leq\lambda<(2\pi)/\eta$. In this case the integral
$$\phi(\lambda)=\phi(0)+u^0(0)\,\Omega_G\int\limits_0^\lambda\frac{d\lambda}{1+\sin^2\left(\frac{1}{2}
\eta\lambda\right)}$$
leads in analogy to (\ref{rechnung1}) under the restriction $m(\lambda)=0,1$ to the angular coordinate
$$\phi(\lambda)=\phi(0)+\arctan\left(\sqrt{2}\tan\left(\eta\lambda/2\right)\right)+m(\lambda)\,\pi.$$
We can rewrite this expression in order to allow arbitrary values of $\lambda>0$ which results in
\begin{equation}
\phi(\lambda)=\phi(0)+\arctan\left(\sqrt{2}\tan\left(\eta\lambda/2\right)\right)+\left(m(\lambda)
-m\textstyle{\left(\lambda-\frac{\pi}{\eta}\right)}\right)\pi.
\label{phi2}
\end{equation}
Equations (\ref{radius2}), (\ref{time2}) and (\ref{phi2}) represent the null geodesics for our special initial conditions.

\end{document}